\documentclass[twocolumn,aps,pra,superscriptaddress]{revtex4-1}  

\usepackage{amsmath,amsthm,graphicx}
\usepackage{dsfont}
\usepackage{graphics}
\DeclareMathOperator{\Tr}{Tr}
\usepackage{color}
\begin{document}
 
\newcommand{\bra}[1]{\langle #1|}

\newcommand{\ket}[1]{|#1\rangle}

\newcommand{\braket}[2]{\langle #1|#2\rangle}

\title{Optimal discrimination of single-qubit mixed states}
\author{Graeme Weir}
\email{g.weir.2@research.gla.ac.uk}
\affiliation{School of Physics and Astronomy, University of Glasgow, Glasgow G12 8QQ, United Kingdom}
\author{Stephen M. Barnett}
\affiliation{School of Physics and Astronomy, University of Glasgow, Glasgow G12 8QQ, United Kingdom}
\author{Sarah Croke}
\affiliation{School of Physics and Astronomy, University of Glasgow, Glasgow G12 8QQ, United Kingdom}

\begin{abstract}
We consider the problem of minimum-error quantum state discrimination for single-qubit mixed states. We present a method which uses the Helstrom conditions constructively and analytically; this algebraic approach is complementary to existing geometric methods, and solves the problem for any number of arbitrary signal states with arbitrary prior probabilities. It has long been known that the minimum error probability is given by the trace of the Lagrange operator, $\Gamma$. The remarkable feature of our approach is the central role played not by $\Gamma$, but by its inverse.

\end{abstract}

\maketitle

\section{Introduction}
A fundamental task in quantum theory, and quantum information in particular, is that of distinguishing between different possible preparation procedures: quantum state discrimination. In quantum theory measurement is an active, rather than a passive process: there are unavoidable restrictions on which observables may be jointly measured, and generically, measurement causes disturbance. The limits to state discrimination thus are inextricably linked with the ultimate limits of measurement in quantum theory. The problem is thus ubiquitous in the field of quantum information and indeed more broadly, with applications including, for example, discrimination of operations in quantum metrology \cite{bae2015quantum}, bounding the dimension of a system Hilbert space given incomplete information \cite{brunner2013dimension}, sharing information through imperfect cloning \cite{buvzek1996quantum,bruss1998optimal,bruss2000phase,keyl1999optimal}, and others \cite{bae2015quantum}.

The problem of quantum state discrimination is perhaps most naturally expressed as a problem in quantum communications, in which it is stated as follows: the quantum states $\{\rho_j\}$ make up an ``alphabet" which can be used to craft a message. A sender, Alice, sends a message to a receiver, Bob, in which each $\rho_j$ has some \emph{a priori} probability $p_j$ of being sent. Unless the signal states are mutually orthogonal, there will be some error in any attempt to determine which state was sent. If Bob knows the set of signal states and their probabilities, what measurement strategy should he use in order to optimally decipher Alice's message?

The study of quantum state discrimination has a long history, beginning with the pioneering work of Helstrom, Holevo, and others in the 1960s and 1970s \cite{helstrom1976quantum,holevo1978asymptotically}, who sought to understand the fundamental limits imposed by quantum theory on optical communications. Since then there has been much further development, with the construction of strategies based on various figures of merit \cite{ivanovic1987differentiate,dieks1988overlap,peres1988differentiate,jaeger1995optimal,chefles1998unambiguous,chefles2001unambiguous,sasaki1999accessible,chitambar2013revisiting,croke2006maximum,barnett1997experimental,andersson2002minimum,chou2004minimum,ha2013complete,bae2013minimum,rosati,bruss1998optimal,bruss2000phase,keyl1999optimal,yuen1975optimum,helstrom1976quantum,holevo2011probabilistic}, as well as ones that interpolate between these \cite{herzog2015optimal,fiuravsek2003optimal}. We are interested here in the minimum error strategy, based on perhaps the most natural figure of merit: what is the highest possible probability of correctly identifying the state?

Necessary and sufficient conditions which an optimal measurement must satisfy are known \cite{helstrom1976quantum}, and further, the problem may be cast as a semi-definite program, which may be solved efficiently numerically \cite{jevzek2002finding}. If an analytical solution is desired, these conditions are most naturally used to check optimality of a candidate measurement procedure, but are not constructive. For some time therefore almost all known analytic solutions were in special cases in which a symmetry property of the states could be used to guide the design of optimal measurements \cite{andersson2002minimum,mochon2006family,ban1997optimum,chou2004minimum}. Some progress has since been made, however, in employing these conditions to construct solutions, and in the qubit case in particular, the Bloch sphere representation proves useful in providing a geometric picture \cite{hunter2004results,deconinck2010qubit,bae2013minimum,ha2013complete,samsonov2009minimum}. Methods of constructing optimal measurements were given first for equi-probable pure qubit states \cite{hunter2004results}, and later for mixed states and for arbitrary prior probabilities \cite{bae2013minimum,ha2013complete,ha2014discriminating,deconinck2010qubit}, the latter using a geometric approach.

In this paper, we give an alternative method of constructing optimal measurements from the minimum error conditions. Previous work has demonstrated that finding a single operator $\Gamma$, sometimes referred to as the Lagrange operator, is equivalent to solving the minimum error discrimination problem: the trace of this operator gives the optimum probability of success, and optimal measurements may be readily constructed once it is known. We construct linear constraints on this operator and its inverse, which in the qubit case may be readily solved for $\Gamma$ and thereby the optimal measurement. Our algebraic approach is complementary to the geometric approach presented in \cite{bae2013minimum}, the results of which were applied to explicitly calculate the optimal probability of correct discrimination for the case of three mixed qubit states in \cite{ha2013complete}. In contrast to the existing approaches, our method reduces the problem of state discrimination to the simpler one of solving a series of linear equations.

\section{The minimum-error conditions}
We begin by introducing the minimum-error conditions \cite{helstrom1976quantum}. We wish to distinguish between states $\{ \rho_i \}$, with \emph{a priori} probabilities $p_i$. Any physically allowed measurement may be described mathematically as a probability operator measure (POM) \cite{barnett2009quantum}, also known as a positive operator-valued measure (POVM) \cite{peres2006quantum}, that is a set of operators $\{ \pi_i \}$ satisfying:
\begin{eqnarray*}
\pi_i & \geq & 0 \\
\sum_i \pi_i &=& \mathds{1}.
\end{eqnarray*}
The Born rule, expressing the probability of obtaining outcome $j$ in a measurement on a system prepared in state $\rho$ is given by:
\begin{equation}
{\rm P}(j|\rho) = {\rm Tr}(\rho \pi_j),
\end{equation}
and the above conditions ensure that these probabilities are positive and sum to 1. A ``click" at the detector corresponding to element $\pi_j$ is taken to indicate that the state $\rho_j$ was sent. Bob's probability of correctly guessing the state Alice sent is then given by ${\rm P}_{\rm corr}=\sum_{i=0}^{n-1}p_i\Tr(\pi_i\rho_i)$, and it is this that we wish to maximise in the minimum error problem (Bob's probability of error, of course, is given by $1-{\rm P}_{\rm corr}$). Clearly to distinguish between $n$ states we therefore need a measurement with at most $n$ outcomes. The number of outcomes may be less than $n$ (or equivalently, some of the operators $\pi_i$ may be zero) if Bob's measurement procedure is such that some states are never identified. Indeed in some cases the optimal measurement is simply to always guess the most likely state \cite{hunter2003measurement} and never identify any other state. 

The solution to the problem of minimum-error quantum state discrimination is equivalent to finding a POVM satisfying the conditions \cite{helstrom1976quantum,holevo2011probabilistic,yuen1975optimum}:
\begin{eqnarray}
\Gamma-p_j\rho_j &\geq & 0 \quad \forall j, \label{H1} \\
\pi_i(p_i\rho_i-p_j\rho_j)\pi_j &=& 0 \quad \forall i, j, \label{H2}
\end{eqnarray}
where $\Gamma=\sum_i p_i \rho_i \pi_i$. The first condition is both necessary and sufficient for $\{ \pi_i \}$ to describe an optimal measurement procedure, and we note that the conditions are not independent; the second, which is necessary but not sufficient, follows from the first. It is useful, however, to give both conditions, because often the second is more convenient to use in practice. Note that $\Gamma$ is a Hermitian operator $\Gamma = \Gamma^\dagger = \sum_i p_i \pi_i \rho_i$, which follows from condition (\ref{H1}), and may be seen explicitly by summing over both $i$ and $j$ in condition (\ref{H2}). An alternative condition is obtained by summing over $i$ in eqn. (\ref{H2}), giving:
\begin{equation}
\label{H3}
\left( \Gamma - p_j \rho_j \right) \pi_j = 0.
\end{equation}
This is a necessary (but not sufficient) condition on any optimal measurement $\{ \pi_j \}$, and is central to our and other methods \cite{bae2013minimum,ha2013complete}, allowing us to construct operators $\pi_j$ satisfying $\Gamma = \sum_i p_i \rho_i \pi_i$ once a candidate $\Gamma$ is given. Indeed, both $\pi_j$ and $\Gamma - p_j \rho_j$ [according to inequality (\ref{H1})] are positive operators, and thus eqn. (\ref{H3}) can hold only if they are orthogonal, that is, $\pi_j$ is entirely within the kernel (or the eigensubspace corresponding to zero eigenvalue) of $\Gamma - p_j \rho_j$. It follows that $\pi_j$ can be non-zero only if $\Gamma - p_j \rho_j$ has at least one zero eigenvalue. We further note that
\begin{equation}
\Tr(\Gamma)=\sum_{i=0}^{n-1}p_i\Tr(\pi_i\rho_i)= {\rm P}_{\rm corr}.
\end{equation}
Therefore if we can find $\Gamma$, we find both the optimal probability of success, and a way of constructing the optimal measurement operators. The problem of finding the optimal measurement $\{ \pi_i \}$, a set of $n$ operators, is thus equivalent to finding a single positive operator $\Gamma$ satisfying the condition (\ref{H1}) and from which operators $\{ \pi_j \}$ satisfying (\ref{H3}) and forming a POVM may be constructed. Indeed the so-called dual problem in the semi-definite programming approach consists of finding the operator $\Gamma$ with minimum trace that satisfies condition (\ref{H1}) for all $j$. Further, as is stressed by Bae \cite{bae2013structure}, the operator $\Gamma$ is unique for a given set of states, while the optimal measurement may not be - for example, in the case of $N\geq4$ symmetric states \cite{hunter2004results,mochon2006family,sasaki1999accessible}.

\section{Qubit state discrimination}
There has recently been much progress in using the Helstrom conditions constructively, to find optimal measurements, with particular success in the qubit case. Hunter was perhaps the first to attempt this in a systematic way, showing how to construct optimal measurements for all sets of equi-probable pure states \cite{hunter2004results}. Samsonov later presented an algorithmic solution to the Helstrom conditions for pure qubits with arbitrary prior probabilities \cite{samsonov2009minimum}. Deconinck and Terhal gave an elegant geometric interpretation of the Lagrange operator $\Gamma$ in the Bloch sphere picture as the minimum enclosing ball of a suitably defined set of balls. A similar interpretation can be given in higher dimensions \cite{tysonremark}. This gives rise to a linear time algorithm for a set of $N$ arbitrary qubit states with arbitrary probabilities. Recently Bae used the so-called Karush-Kuhn-Tucker (KKT) conditions from semi-definite programming to define complementary states $\{ \sigma_j \}$ with weights $r_j$ such that
$$
\Gamma = p_i \rho_i + r_i \sigma_i = p_j \rho_j + r_j \sigma_j.
$$
The geometric structure of the complementary states $\sigma_j$ may be deduced from the conditions and the geometric structure of the signal states $\rho_j$, and in turn used to construct $\Gamma$. Bae discusses the qubit case, in which the Bloch sphere provides a convenient geometric picture, and the full details for three mixed qubit states were later calculated by Ha and Kwon \cite{ha2013complete}.

We begin with some general considerations concerning the qubit state discrimination problem, and then discuss our method, which constructs $\Gamma$ directly, without reference to complementary states. Firstly, we note that for each $j$ the operator $\Gamma - p_j \rho_j$ can have two, one, or no zero eigenvalues, corresponding to the zero operator, a rank-one operator, and a positive-definite operator respectively:
\begin{enumerate}
\item If $\Gamma - p_j \rho_j = 0$ for some $j$, then $\Gamma = p_j \rho_j$, which can only hold if $p_j \rho_j - p_k \rho_k \geq 0$ for all $k$. The no measurement strategy is then an optimal measurement, $\pi_k = I \delta_{jk}$ \cite{hunter2003measurement}.
\item $\Gamma - p_j \rho_j$ has a single zero eigenvalue. If $\pi_j$ is non-zero, it is a weighted projector onto the corresponding eigenstate.
\item If $\Gamma - p_j \rho_j$ is positive definite (all eigenvalues strictly greater than zero), then according to condition (\ref{H3}) it follows that $\pi_j = 0$ for every optimal measurement, and the corresponding state is never identified.
\end{enumerate}
Given a set of qubit states $\{ \rho_j \}$ with \emph{a priori} probabilities $p_j$, it is easily checked whether for some $j$
\begin{equation}
p_j \rho_j - p_k \rho_k \geq 0, \quad \forall k.
\label{nomeas}
\end{equation}
If this does hold for some $j$, the optimal strategy is not to measure at all and simply guess $\rho_j$. For all other ensembles, it follows that the optimal measurement is made up of rank-one weighted projectors,
$$
\pi_j = c_j \ket{\phi_j} \bra{\phi_j}
$$
for some $c_j$ satisfying $0 \leq c_j \leq 1$, and where $\ket{\phi_j}$ is the eigenstate of $\Gamma - p_j \rho_j$ corresponding to the zero eigenvalue. Note that this is completely general for qubits, and holds whether $\rho_j$ are pure or mixed states. Thus, for an optimal measurement each operator $\pi_j$ is uniquely defined, up to a multiplying factor. There may however be more than one way of choosing the coefficients $c_j$ such that the $\pi_j$ thus found sum to the identity.

Secondly, we note that for minimum error discrimination of an arbitrary set of qubit states there always exists an optimal measurement with at most four outcomes. Intuitively, the constraint $\sum_i\pi_i=\mathds{1}$ contains only $d^2$ independent linear constraints, where $d$ is the dimension of our space: if a set of $N>d^2$ elements $\{\pi_j\}$ satisfies this, there is always a subset of these which, when appropriately weighted, also forms a resolution of the identity. A measurement with $>d^2$ outcomes can always be decomposed as a probabilistic mixture of measurements with at most $d^2$ outcomes. If the mixture results in an optimal procedure, then any of the component measurements must also be optimal \cite{hunter2004results,deconinck2010qubit}. A more complete proof of this may be found in the literature, e.g., \cite{chiribella2007continuous}.

Finally, note that the number of outcomes in our optimal measurement corresponds to the number of states that are identified with non-zero probability by the measurement: additional states are never identified. Denoting the number of outcomes $k$, the cases $k=1$ and $k=2$ are well-known, as these correspond to the no-measurement strategy and the Helstrom two-state discrimination measurement respectively \cite{helstrom1976quantum,barnettcroke2009quantum}. In the cases $k=3$ and $k=4$ it is more difficult to find optimal measurements, although as discussed above, strategies for these cases have been recently suggested.

For qubits, the Pauli operators together with the identity form a convenient basis in which to express any operator on the space. Thus, for example, we can write
\begin{equation}
\label{Gamma}
\Gamma=\frac{1}{2}(a\mathds{1}+\vec{b}\cdot\vec{\sigma}),
\end{equation}
where $a>0$, $\vec{b}$ is a real three-dimensional vector, and $\vec{\sigma}$ is the vector of Pauli operators: $\vec{\sigma} = (\sigma_x, \sigma_y, \sigma_z)$. It will be convenient in what follows to also use such a representation for the inverse $\Gamma^{-1}$, and it is easily verified that:
\begin{equation}
\Gamma^{-1}=\frac{2}{a^2-|b|^2}(a\mathds{1}-\vec{b}\cdot\vec{\sigma}).
\end{equation}
Note that $\Gamma$ is a strictly positive operator in the space spanned by the states to be discriminated, and so the inverse is always well-defined, as is the square-root, which we will use later. We thus need four parameters to completely specify $\Gamma$, and we discuss now how to construct four constraints, which are readily inverted to construct $\Gamma$.

Suppose there is an optimal measurement which identifies $k>2$ states: we will show that for each of these we may obtain one constraint on the parameters of $\Gamma$. It is of course not obvious \emph{a priori} which states will be identified by an optimal measurement; however, we can construct a candidate $\Gamma$, under the assumption that a particular subset of our states are identified in an optimal measurement, and then verify that this results in a physically allowed measurement procedure. We will return to this later. According to the discussion above therefore, for each of these $k$ states the operator $\Gamma - p_j \rho_j$ has a single zero eigenvalue. Let us consider first the pure state case: $\rho_j = \ket{\psi_j} \bra{\psi_j}$. If we pre- and post-multiply condition (\ref{H1}) by $\Gamma^{-1/2}$ we find:
$$
\mathds{1} - p_j \Gamma^{-1/2} \ket{\psi_j} \bra{\psi_j} \Gamma^{-1/2} \geq 0.
$$
We further require that this operator has exactly one zero eigenvalue, which in turn requires that the second term has modulus 1:
\begin{equation}\label{gammatrick}
p_j \bra{\psi_j}\Gamma^{-1}\ket{\psi_j}= 1.
\end{equation}
A similar relation was pointed out by Mochon \cite{mochon2006family}, who discussed the inverse problem of characterising the sets of states and corresponding probabilities for which a given measurement procedure was optimal, although it does not seem to have been used constructively in the literature. Thus we find
\begin{equation}
2p_j\bra{\psi_j}(a\mathds{1}-\vec{b}\hat{\sigma})\ket{\psi_j}=a^2-|b|^2.
\end{equation}
Alternatively, if $\rho_j = \ket{\psi_j} \bra{\psi_j}$ has Bloch vector $\hat{r}_j$ (a unit vector as $\rho_j$ is a pure state), $\rho_j = \frac{1}{2} \left( \mathds{1} + \hat{r}_j \cdot \vec{\sigma} \right)$, we may write:
\begin{equation}
2 p_j \left( a - \hat{r}_j \cdot \vec{b} \right)=a^2-|b|^2.
\label{C1}
\end{equation}
Each state gives rise to one such constraint, resulting in $k$ independent constraints on $\Gamma$.

This procedure is readily adapted to apply also to mixed states for the qubit case. Note that for qubit states, every mixed state can be written as a mixture of a pure state and the identity: $\rho_j = \alpha_j \ket{\psi_j} \bra{\psi_j} + \beta_j \frac{1}{2} \mathds{1}$, where $\alpha_j + \beta_j = 1$. The requirement that $\Gamma - p_j \rho_j \geq 0$ then becomes:
$$
\Gamma - \frac{1}{2} p_j \beta_j \mathds{1} - p_j \alpha_j \ket{\psi_j} \bra{\psi_j} \geq 0,
$$
and using the same reasoning as previously, if we require this operator have exactly one zero eigenvalue we obtain:
$$
p_j \alpha_j \bra{\psi_j} \left( \Gamma - \frac{1}{2} p_j \beta_j \mathds{1} \right)^{-1} \ket{\psi_j} = 1.
$$
Explicitly, this gives:
\begin{align}
&2p_j \alpha_j \bra{\psi_j}[(a-p_j\beta_j)\mathds{1}-\vec{b}\cdot \vec{\sigma}]\ket{\psi_j}\nonumber \\
&=(a-p_j\beta_j)^2-|b|^2,
\end{align}
and after a litte rearranging, again writing $\ket{\psi_j} \bra{\psi_j} = \frac{1}{2} \left( \mathds{1} + \hat{r}_j \cdot \vec{\sigma} \right)$, we find
\begin{equation}
2p_j \left( a - \alpha_j \hat{r}_j \cdot \vec{b} - p_j\beta_j (\alpha_j + \frac{1}{2} \beta_j) \right) =a^2-|b|^2.
\label{C2}
\end{equation}
If there are $k$ states identified by the optimal measurement this procedure, in both the pure state and mixed state case, gives $k$ equations for the parameters of $\Gamma$. Clearly if $k=4$ this is enough to construct $\Gamma$. We further note that in equations (\ref{C1}) and (\ref{C2}) the non-linear right hand side is independent of $j$, thus we can easily take linear combinations to obtain $k-1$ linear equations. For $k=4$ these are readily solved to write all parameters in terms of a single one, e.g. $a$, which is finally determined by solving one quadratic equation.

Thus we can construct optimal measurements with $k=1,2,$ or $4$ outcomes. For $k=3$ we do not yet have enough constraints to determine $\Gamma$; a further constraint, however, is readily constructed, as we now discuss. We first note that for the special case in which three signal states lie in an equatorial plane of the Bloch sphere (as in \cite{andersson2002minimum,myfirstpaper}), we know from symmetry that the POVM elements, and therefore also $\Gamma$, must lie in the same plane as the signal states, thus giving us our final constraint. More generally, for the case of three equiprobable pure qubit states it is always possible to choose a representation in which the states sit at the same latitude of the Bloch sphere. The optimal measurement operators $\pi_j$ then lie in the equator of the sphere, and $\Gamma$ has the same latitude as the signal states \cite{hunter2004results}.

We can generalize this idea to both pure and mixed states, and to non-equal prior probabilities. We first note that for a three outcome measurement, all three elements of the POVM must lie in an equatorial plane of the Bloch sphere in order to form a resolution of the identity. Without loss of generality we choose our axes so that this is the $z=0$ plane. That is, we can always choose our axes so that $\pi_j = \frac{1}{2} \left( c_j \mathds{1} + \vec{d_j}\cdot\vec{\sigma} \right) $, with $d_{jz} = 0$, $\forall j$. Referring now to condition \ref{H3}, it follows that
$$
\Gamma-p_j\rho_j \propto \frac{1}{2} \left( c_j \mathds{1} - \vec{d_j}\cdot\vec{\sigma} \right),
$$
and thus $\langle \Gamma \sigma_z \rangle - p_j \langle \rho_j \sigma_z \rangle = 0$. Finally we therefore require
$$
b_z = p_j \langle \rho_j \sigma_z \rangle = p_j \alpha_j \hat{r}_{jz}, \quad \forall \quad j,
$$
where as before $\rho_j = \frac{1}{2} \left( \mathds{1} + \alpha_j \hat{r}_j \cdot\vec{\sigma} \right)$. Thus if we define our $z$-axis to be such that the $z$-component of $p_j\rho_j$ is the same for each of the three signal states identified, then $\Gamma$ also has the same $z$-component, and the optimal measurement operators lie in the equatorial plane. Note that a similar discussion may be found in \cite{herzog2015optimal}.

Thus for a given set of qubit states $\{ \rho_i \}$ with arbitrary priors $\{ p_i \}$, if there exists an optimal minimum error measurement which identifies a subset of $k=1,2,3,4$ of these states, we have shown how to construct $\Gamma$, which in turn allows us to construct the optimal measurement. We illustrate below in an example how this may be employed in practice to find optimal measurements, and discuss later the problem of how we can know in general which states are identified by an optimal measurement.

\section{Example}
To illustrate our method, we consider the problem of discriminating between three pure states which are mirror-symmetrically arranged on the equator of the Bloch sphere, previously investigated by Andersson, et. al. \cite{andersson2002minimum}. The states are:
\begin{equation*}
\begin{split}
\ket{\psi_0}=\ket{+}=\frac{1}{\sqrt{2}}(\ket{0}+\ket{1}), \\
\ket{\psi_1}=\frac{1}{\sqrt{2}}(\ket{0}+e^{i\theta}\ket{1}), \\
\ket{\psi_2}=\frac{1}{\sqrt{2}}(\ket{0}+e^{-i\theta}\ket{1}),
\end{split}
\end{equation*}
and these occur with \emph{a priori} probabilities \mbox{$p_0=1-2p$}, $p_1=p_2=p$, with $p\in[0, \frac{1}{2}]$. The so-called trine ensemble occurs at $\theta=\frac{2\pi}{3}$ \cite{myfirstpaper}.

We begin by noting that as the states are all pure it is not possible to satisfy conditions (\ref{nomeas}) and the no-measurement solution is never optimal. We next check to see when a two outcome measurement is optimal. Note that due to the symmetry the only sensible two-outcome measurement is one distinguishing $\ket{\psi_1}$ and $\ket{\psi_2}$: the optimal such measurement is a projective measurement in the eigenbasis of $\sigma_y$; $\pi_i = \ket{\phi_i} \bra{\phi_i}$, where $\ket{\phi_1}=\frac{1}{\sqrt{2}}(\ket{0}+i\ket{1})$ and $\ket{\phi_2}=\frac{1}{\sqrt{2}}(\ket{0}-i\ket{1})$. It is straight-forward to calculate $\Gamma_{\rm 2-element}$, and we note that condition \ref{H1} is satisfied for $j=1,2$ by construction. In order to check this condition for $j=0$, it is enough to verify that ${\det{(\Gamma_{\rm 2-element}-p_0\rho_0)}\geq0}$. We find as in \cite{andersson2002minimum}, that this holds when
\begin{equation}
p\geq\frac{1}{2+\cos(\frac{\theta}{2})[\cos(\frac{\theta}{2})+\sin(\frac{\theta}{2})]}.
\label{pcondition}
\end{equation}
The corresponding optimal probability of correctly identifying the state is given by
\begin{equation*}
{\rm Tr}(\Gamma_{2-element})=p(1+\sin\theta).
\end{equation*}
When condition \ref{pcondition} does not hold, we know that a three outcome measurement is optimal, and can use the method outlined above to find this. We first note that $\hat{r}_z=0$ for each of our signal states. Thus, as discussed above, $\Gamma$ must also have $b_z=0$, and lies in the equatorial plane. Further, using equation \ref{C1}, we obtain the following three constraints on $a, b_x$, and $b_y$:
\begin{align*}
a^2-|b|^2&=2(1-2p)(a+b_x)\\
a^2-|b|^2&=2p(a+b_x\cos\theta+b_y\sin\theta)\\
a^2-|b|^2&=2p(a+b_x\cos\theta-b_y\sin\theta)
\end{align*}
It is clear from the latter two that $b_y=0$. The remaining equations are readily solved for $a$ and $b_x$, giving:
\begin{align*}
\begin{split}
a&=b_x\frac{p\sin^2\frac{\theta}{2}+1-2p-p\cos^2\frac{\theta}{2}}{3p-1}\\
b_x&=\frac{(3p-1)(1-2p)}{1-2p-p\cos^2\frac{\theta}{2}}
\end{split}
\end{align*}
The corresponding probability of correctly identifying the state ${\rm P}_{\rm Corr}$ is then given by:
\begin{align*}
{\rm P}_{\rm Corr} &= {\rm Tr}(\Gamma) = a \\
&=\frac{(1-2p)(p\sin^2\frac{\theta}{2}+1-2p-p\cos^2\frac{\theta}{2})}{1-2p-p\cos^2\frac{\theta}{2}}
\end{align*}
which agrees with the solution provided in \cite{andersson2002minimum}.

\begin{figure}
\centering
\includegraphics[scale=0.77]{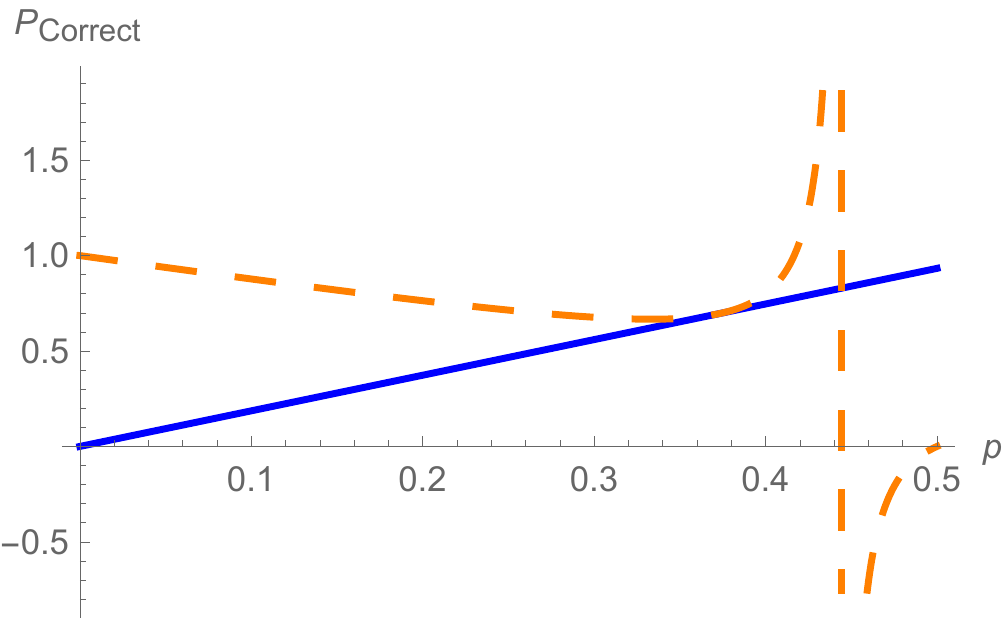}
\caption{The two functions we obtained for ${\rm P}_{\rm Corr}$ plotted against $p$ for the optimal two-element (solid blue line) and three-element (dashed orange line) POVMs - we can see that for $p>0.373$, the three-element POVM appears to be superior to the two-element POVM. However, this turns out to no longer be physically realisable, and fails to satisfy the condition in equation \ref{H1}. Note that the function we find for ${\rm Tr}(\Gamma)$ is not strictly positive - at no point have we assumed that $\Gamma$ must be positive.}
\label{PCorr}
\end{figure}

We finally note that we found the region in which a three outcome measurement was necessary by first finding the region in which a two-outcome measurement was optimal. If we use our method to find a candidate $\Gamma$ in the region where in fact the optimal measurement has only two outcomes, we find that even though it is possible to construct $\Gamma$, it is not possible to construct a physically allowed measurement from the conditions \ref{H3}, and the method fails. Further, it can sometimes return probabilities that are greater than 1, clearly indicating that something has gone wrong. This is illustrated in Fig \ref{PCorr}.

\section{Discussion}
We have presented a method to construct optimal minimum error measurements from the known necessary and sufficient conditions. If we know which of a set of states are identified by an optimal measurement, the method presented here allows us to construct four linear conditions on either $\Gamma$ or $\Gamma^{-1}$, from which we have enough information to reconstruct $\Gamma$. The remaining problem we have not addressed, and which is common to other methods in the literature \cite{herzog2015optimal,ha2013complete}, is how to find which states our measurement should identify. We finish with some comments on this problem.

In the worst case, we can find the optimal measurement by exhaustive search: we first check if the no measurement solution is optimal. If yes then we are done, and if not then we know that $k>1$. We then check whether any measurement identifying just two of the states is optimal. This consists of constructing optimal measurements for each pair of states, and checking the condition (\ref{H1}) for the remaining $N-2$ states in each case. There are $ \binom{N}{2} $ such measurements. If none of these are optimal, then we know $k>2$, and so on. This requires constructing $\sum_{k=1}^4 \binom{N}{k}$ [i.e. $O(N^4)$] candidate $\Gamma$ operators, and for each one checking $O (N)$ conditions, thus we require $O(N^5)$ operations, in the worst case. Our detailed results for the case of three symmetric states with arbitrary priors, which we discuss elsewhere \cite{myfirstpaper}, indicate that for almost all prior probabilities the optimal measurement has only two outcomes. Thus we expect that in many cases an optimal measurement will be found faster than $O (N^5)$ operations.

For a given set of states, the method we present here allows us to characterise the entire parameter space of prior probabilities, beginning with the no-measurement solution, through those regions in which a two-outcome measurement is optimal, and constructing three- and then four-outcome solutions for the remaining regions, as shown in the example above. We note also that for specific cases numerical methods can also be used to determine which states are identified by an optimal measurement, and once this is known our method may be used to find an exact analytical solution for the optimal probability of correctly identifying the state and to find optimal measurements.

We have introduced an analytical method, complementary to the geometric approach in the literature, for constructing optimal measurements for minimum error state discrimination problems. Our method constructs linear constraints on the so-called Lagrange operator $\Gamma$, and its inverse $\Gamma^{-1}$, from which the optimal $\Gamma$ may readily be found for any qubit state discrimination problem. Although the constraints we present appear elsewhere in the literature in a different context, it seems not to have been recognised that these together give enough information to construct optimal measurements. We have further shown that these are applicable to both pure and mixed states in the qubit case.

In this paper we have discussed the qubit case in detail. We expect that the linear constraints given on $\Gamma$ may also be applied in higher dimensions. The constraints on $\Gamma^{-1}$ may be applied to pure states in higher dimensions, although the mixed state case appears less straight-forward, because it is no longer the case that any mixed state is a mixture of a pure state and the maximally mixed states. We leave a full discussion of the generalization to higher dimensions for future work.

\textbf{Acknowledgements}

This work was supported by the University of Glasgow College of Science and Engineering (S.C. and G.W.) and by the Royal Society Research Professorships (S.M.B., Grant No RP150122).


\end{document}